\documentclass[preprint,12pt]{elsarticle}

\usepackage{amssymb}

\journal{Nuclear Inst. and Methods in Physics Research, A}

\usepackage{HGTD_TDR-defs}
\usepackage{multirow} 
\usepackage{rotating}
\usepackage{mathrsfs}
\usepackage{booktabs}
\usepackage{longtable}
\usepackage{placeins}
\usepackage{amssymb}
\usepackage{upgreek}
\usepackage{svg}
\usepackage{comment}
\usepackage{pdfpages}
\usepackage{siunitx}
\usepackage{lineno}
\usepackage{float}
\usepackage{lineno}
\usepackage{tikz}
\usepackage[percent]{overpic}
\usepackage{subcaption}
\title{\boldmath Gain suppression study on LGADs at the CENPA tandem accelerator}
 
\author[1]{S.~Braun}
\author[1]{Q.~Buat}
\author[2]{J.~Ding}
\author[1]{P.~Kammel}
\author[2]{S.M.~Mazza}
\author[2]{F.~McKinney-Martinez}
\author[2]{A.~Molnar}
\author[3]{C. Lansdell}
\author[2]{J.~Ott}
\author[2]{A.~Seiden}
\author[2]{B.~Schumm}
\author[2]{Y.~Zhao}

\affiliation[1]{Center for Nuclear Physics and Astrophysics (CENPA), University of Washington, Seattle, WA 98195, USA}
\affiliation[2]{SCIPP, University of California Santa Cruz, 1156 High Street, Santa Cruz (CA), US}
\affiliation[3]{West-Virginia University, Morgantown, WV 26506, USA}

\usepackage{sfmath}

\begin{document}

\begin{frontmatter}

\begin{abstract}
Low-Gain Avalanche Detectors (LGADs) are a type of thin silicon detector with a highly doped gain layer that provides moderate internal signal amplification. One recent challenge in the use of LGADs, studied by several research groups, is the gain suppression mechanism for large localized charge deposits. Using the CENPA Tandem accelerator at the University of Washington, the response of the LGADs to MeV-range energy deposits from a proton beam was studied. Two LGAD prototypes and a PIN diode were characterized, and the gain of the devices was determined as a function of bias voltage, incidence beam angle and proton energy.
This study was conducted in the scope of the PIONEER experiment, an experiment proposed at the Paul Scherrer Institute to perform high-precision measurements of rare pion decays.
A range of deposited charge from Minimum Ionizing Particle (MIP, few 10s of KeV) from positrons to several MeV from the stopping pions/muons is expected in PIONEER; the detection and separation of close-by hits in such a wide dynamic range will be a main challenge of the experiment. To achieve this goal, the gain suppression mechanism has to be understood fully.
\end{abstract}


\end{frontmatter}


\section{Introduction}
\label{sec:intro}

Low Gain Avalanche Detectors (LGADs) are thin silicon detectors with internal gain~\cite{bib:LGAD} capable of providing measurements of minimum-ionizing particles with a time resolution as good as 17 ps~\cite{Zhao:2018qkg}.
This makes LGADs the prime candidate sensor technology for achieving 4D tracking in future experiments. 
Furthermore, the fast rise time and short full charge collection time (as low as 1~ns) of LGADs are suitable for high repetition rate measurements in photon science and other fields.
LGADs are the chosen technology for near-future large-scale applications like the timing layer upgrade for HL-LHC of the ATLAS~\cite{CERN-LHCC-2020-007} and CMS~\cite{CMS:2667167} experiments at CERN, the EPIC detector at the Electron-Ion Collider at BNL~\cite{EIC} and the PIONEER experiment~\cite{Mazza:2021adt}.
This study was of primary interest to the PIONEER collaboration since LGAD energy linearity is particularly important for particle identification in the inner active target detector (ATAR)~\cite{Mazza:2021adt}.

The mechanism underlying gain suppression or saturation, i.e., the loss of gain with increasing signal charge, has been studied in recent years by several research groups~\cite{s22031080,Curr_s_2022}.
The saturation is triggered by the shielding of the electric field in the gain layer caused by the multiplication of the charge carriers in the bulk, and increases with the density of the deposited charge. 
It depends on the total amount of initial deposited charge, the sensor gain (which, in LGADs, is proportional to the applied bias voltage), and the angle of incidence with respect to the charge drift direction towards the gain layer.
This is because the charge carrier drift is more distributed across the gain layer for an angled track, reducing the saturation effect. 
This effect also depends on the initial gain (which is the device gain for charge deposition of 1 MIP) of the device; running the device at a higher gain will increase the saturation effect.

\section{Experimental setup}
The Center for Experimental Nuclear Physics and Astrophysics (CENPA) at the University of Washington has a High Voltage Engineering Corporation Model FN tandem Van de Graaff accelerator~\cite{tandem} that is used to perform various accelerator-based experiments.
Protons of 1.8~MeV and 3~MeV momenta were used for this measurement campaign.
The momentum resolution of the proton beam is around 300~ppm, and the beam size is around 2~mm. 

\begin{figure}[h]
    \centering
    \includegraphics[width=0.6\textwidth]{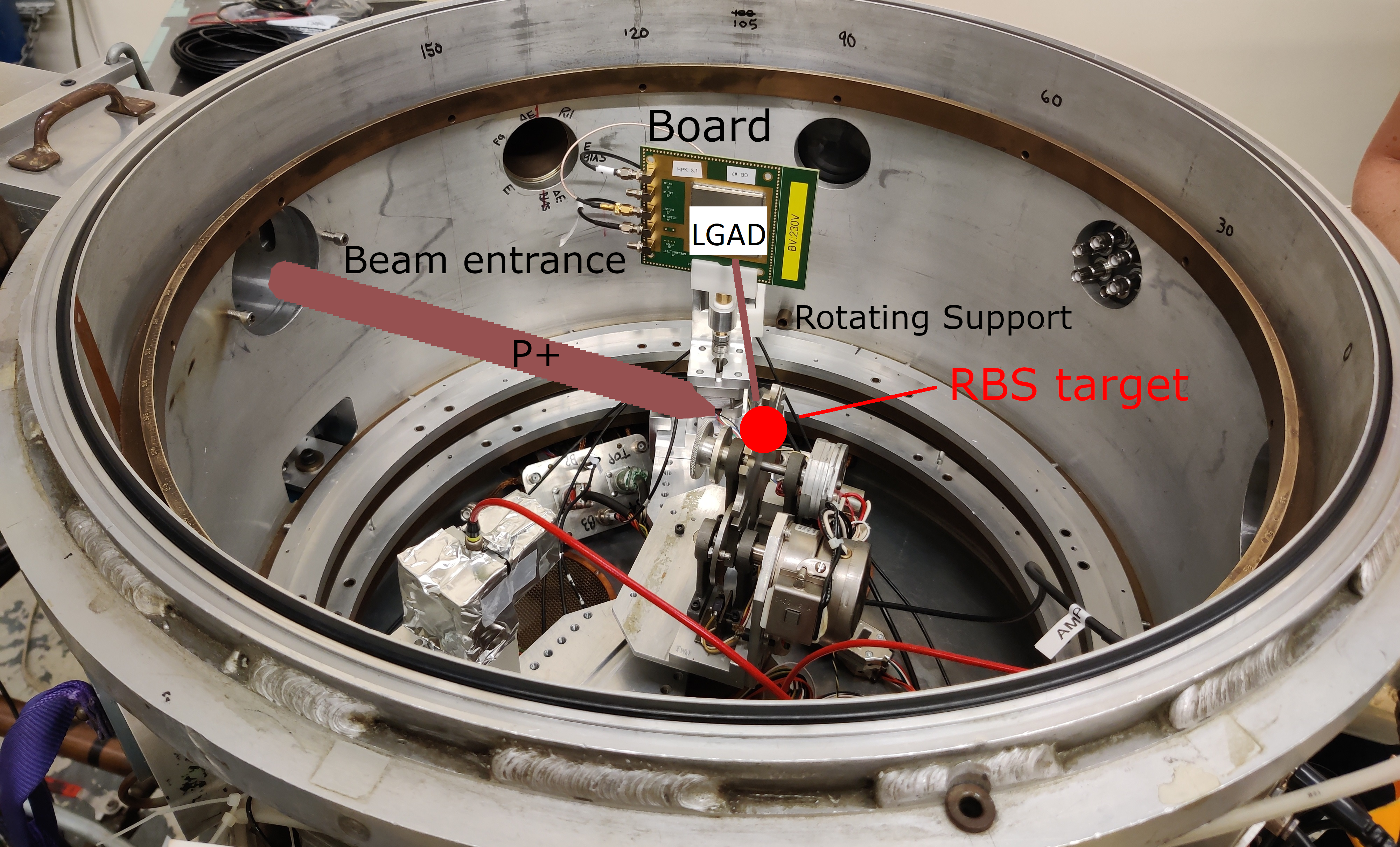}
    \caption{Sensor testing setup with 1-ch board on rotational stage, illustrating the proton beam and backscattered proton beams in a UHV chamber at a CENPA Tandem accelerator beamline.
    }
    \label{fig:setup}
\end{figure}

The tested LGAD sensors are mounted on a fast analog single channel electronic board (around 2~GHz bandwidth) designed at SCIPP~\cite{Padilla_2020} with an additional external box amplifier\footnote{GALI-52+ evaluation board}, the signal is digitized by a 1~GHz bandwidth digital scope\footnote{Tektronix DPO 7104, 8~bit digitizer, 20~Gs, 1~GHz}. Appropriate power supplies provide the amplifiers with the Low Voltages and the sensor with High Voltage bias.
The setup sits in a vacuum chamber directly connected to the beamline, as shown in Fig.~\ref{fig:setup}.
The board is fixed on a rotating support that can be controlled remotely. 
To control the particle incidence rates, the Rutherford Backscattering (RBS) technique was used: the proton beam is scattered off a gold foil target, and the board and its support are mounted on a rotating arm that keeps the board at a 110-degree angle with respect to the incoming beam and points it to the RBS target.
The RBS target is a 110~nm gold foil, providing sufficient particle rate on the sensor under test (around the kHz) while maintaining the energy resolution of the initial beam. Degradation in momentum from the foil used was estimated to be less than 2\% using SIMNRA simulations~\cite{SIMNRA}.
Cables are brought outside with pass-through connectors.

The reference data for beta particles from a $^{90}$Sr source was taken in the SCIPP laboratories with identical sensors and readout boards; the energy of the $^{90}$Sr electron is in the minimum ionizing particle (MIP) range. The system has been previously described in detail in~\cite{Zhao:2018qkg}. Both the CENPA test beam data and the $^{90}$Sr data were analyzed using the same package described in the ``Data analysis'' section.

\section{LGAD sensors}
\label{sec:devices}
\begin{figure}[htbp]
\centering
\begin{subfigure}[b]{0.48\textwidth}
    \includegraphics[width=\textwidth]{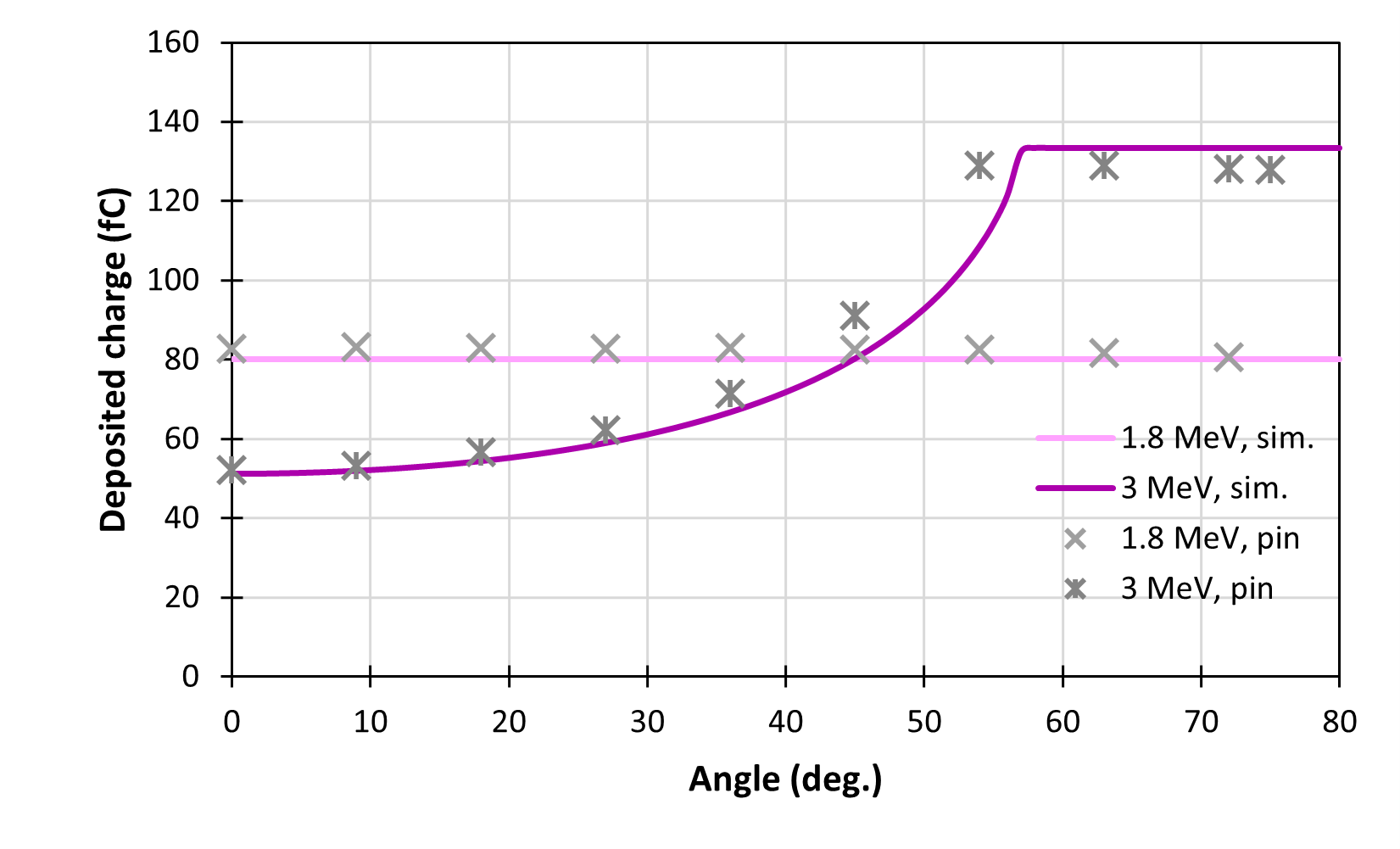}
    \caption{}\label{fig:histo_LGAD}
\end{subfigure}
\begin{subfigure}[b]{0.48\textwidth}
    \includegraphics[width=\textwidth]{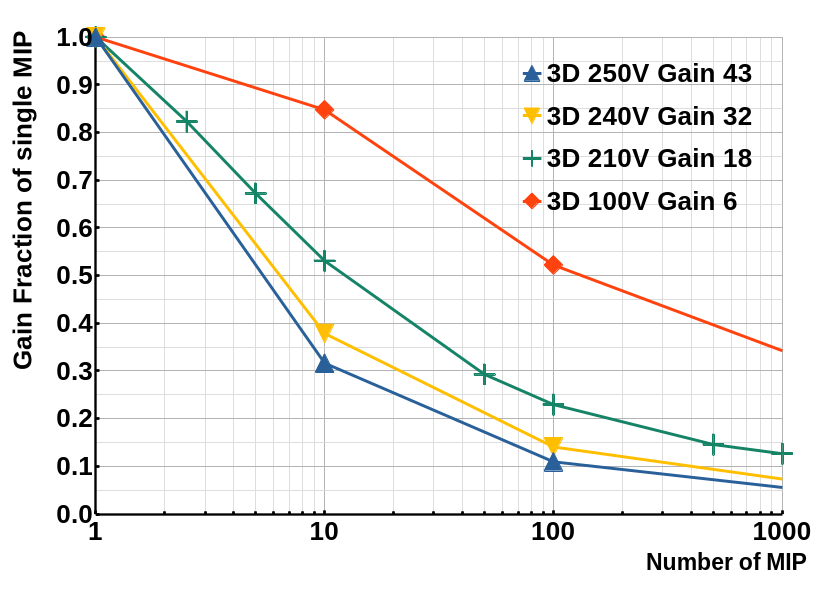}
    \caption{}\label{fig:histo_LGAD}
\end{subfigure}
\caption{(a) SRIM simulation of energy deposited by a proton with initial energy 1.8/3~MeV in 50 µm of Si as a function of incidence angle. In the same plot, the grey points represent the measured deposited charge in a Silicon PIN detector without gain for the protons as a function of the angle. (b) TCAD simulation of gain suppression, as a fraction of gain in respect to 1~MIP deposition, for a simulated Hamamatsu LGAD device (HPK~3.1) for different initial gains at 1~MIP deposition, the charge is injected in a straight line with increased density. 
}
\label{fig:alpha_beta}
\end{figure}

Two standard single pads LGADs and a PIN from HPK (Hamamatsu Photonics), complete characteristics given in~\cite{Padilla_2020}, were tested. The devices tested had \SI{50}{\micro\meter} of active thickness.
A list of the devices' characteristics is shown in Tab.~\ref{tab:LGADs}.
The sensor with a deeper gain layer has an increased gain due to the effects discussed in literature~\cite{Padilla_2020}.
The table shows the breakdown voltage corresponding to the highest possible sensor gain. 
However, during operations inside the vacuum chamber, the maximum applicable bias voltage was reduced by 20-50~V; the cause of this premature breakdown in the vacuum is still under investigation. 
The same devices were tested in a low vacuum chamber beforehand and did not present this behavior.
The high vacuum environment might have caused the reduced breakdown, but there is no definitive proof.
The sensor returned to having a normal breakdown a few months later during laboratory tests.
The simulated and measured charge deposited in the PIN detector for different beam energies and angles is shown in Fig.~\ref{fig:alpha_beta} (a). At 1.8~MeV of energy, the proton always stops (stopping power 120~MeV/cm$^2$g, range in silicon around \SI{40}{\micro\meter}) and deposits about 80~fC of charge; at 3~MeV, the proton initially punches through and stops for an angle of approximately 55~degrees (stopping power 84.33~MeV/cm$^2$g, range in silicon around \SI{90}{\micro\meter}), depositing around 130~fC. Proton stopping power and ranges were calculated with pstar~\cite{pstar}.
In Fig.~\ref{fig:alpha_beta}(b), the relative gain loss simulated with Synopsys Sentaurus TCAD as a function of the number of MIPs injected is shown for the HPK~3.1 device at different initial gains (gain for 1~MIP charge deposition). 
From the simulation, which is qualitative as the effect is not perfectly reproduced by simulation, the gain suppression appears larger for an initial higher gain and increasing charge injection.
To provide a reference, the collected charge for a MIP is around 0.5~fC for the \SI{50}{\micro\meter} thickness of the detector tested (see~\cite{Zhao:2018qkg}, Appendix A); therefore, the injected charge by the proton is around 160 and 260~MIP at 1.8~MeV and 3~MeV, respectively.

\begin{table}[H]
    \centering
    \begin{tabular}{|l|c|c|c|c|c|}
        \hline
         Device & Producer & BV & Thickness & Gain layer & Gain range \\
         \hline
         HPK 3.1 & HPK & \SI{230}{\volt} & \SI{50}{\micro\meter} & shallow & 5-30\\
         HPK 3.2 & HPK & \SI{130}{\volt} & \SI{50}{\micro\meter} & deep & 30-50\\
         HPK PIN & HPK & \SI{400}{\volt} & \SI{50}{\micro\meter} & no gain & 1\\
         \hline
    \end{tabular}
    \caption{List of tested HPK LGADs and PIN devices. All devices have a 1.3$\times$1.3~\si{\milli\meter\squared} single pad geometry.}
    \label{tab:LGADs}
\end{table}

\section{Data analysis}
\label{sec:data_ana}
The 1 GHz digital oscilloscope was used to save the voltage waveforms, which were converted into ROOT files and analyzed offline. The waveforms were imported into a SCIPP data analysis package, which produced event-level variables. 
The waveforms are inverted to show a positive pulse for convenience, and the baseline is corrected using the first flat part of the waveform.
The variables calculated include p-max (pulse maximum), p-min (pulse-minimum), pulse positive area, undershoot area, combined pulse area, rise time, fall time, and the pulse full-width half maximum (FWHM).  
Rise and fall times were calculated as the time interval between the 10\% and 90\% points of the pulse maximum on the rising edge and the 90\% and 10\% of the pulse maximum on the falling edge, respectively. 
The FWHM value was determined as the time interval between the 50\% of the pulse maximum on the rising edge and the 50\% of the pulse maximum on the falling edge.
The pulse positive area was the area from the positive pulse created from the event signal, and the undershoot area was the negative pulse caused by amplifier effects.
The combined area is the addition of the two areas (subtracting the undershoot area from the pulse area), which is then divided by the total amplifier trans-impedance (4700~$\Omega$) to calculate the collected charge.
Lastly, the gain is calculated as the collected charge of the LGAD divided by the collected charge of the same-size PIN with the same environmental running conditions (energy and angle). The PIN performance does not depend on the bias voltage after drift velocity saturation, and the collected charge is the same even before saturation, although with a different pulse shape. Therefore, the bias voltage applied to the PIN does not affect the gain calculation using the collected charge.

\section{Results}
\begin{figure}
    \centering
    \begin{subfigure}[b]{0.48\textwidth}
        \includegraphics[width=\textwidth]{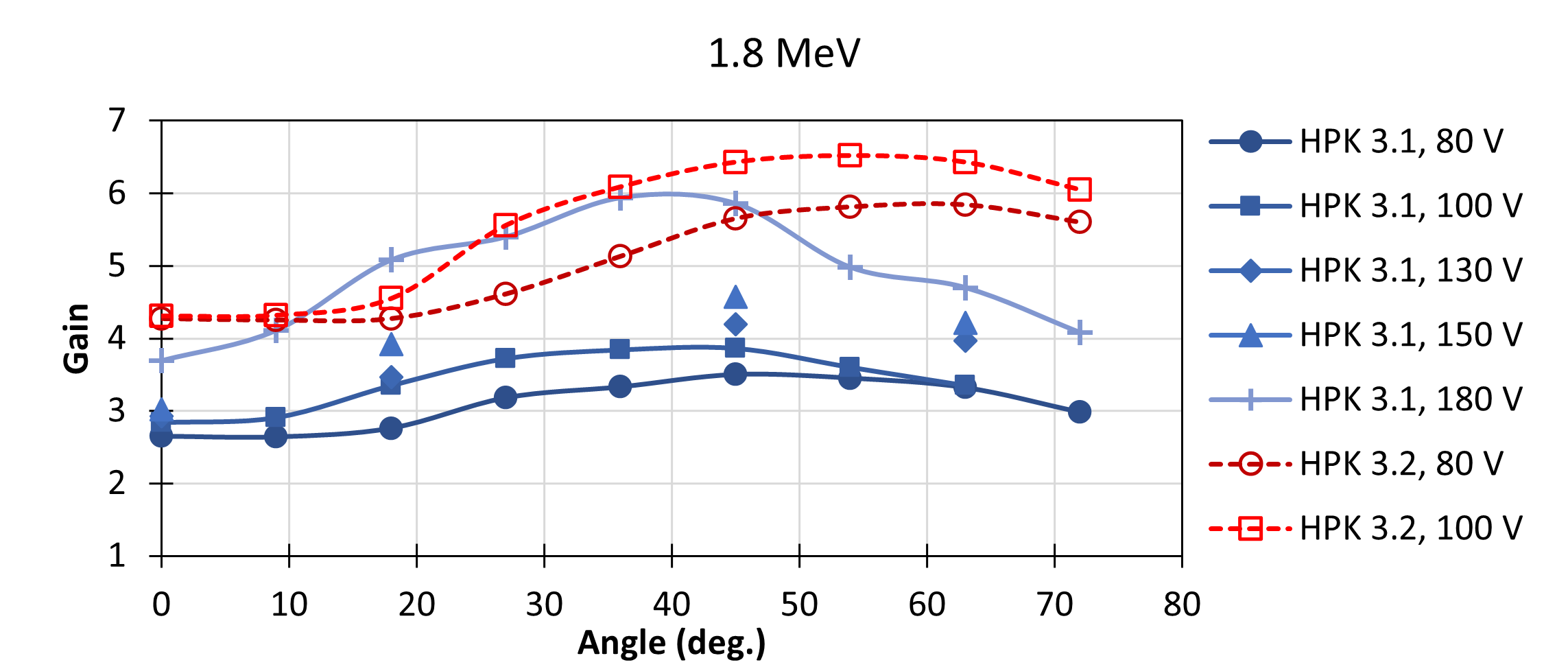}
        \caption{}
        \label{fig:HPK3p1_1p8MeV}
    \end{subfigure}
    \quad
    \begin{subfigure}[b]{0.48\textwidth}
        \includegraphics[width=\textwidth]{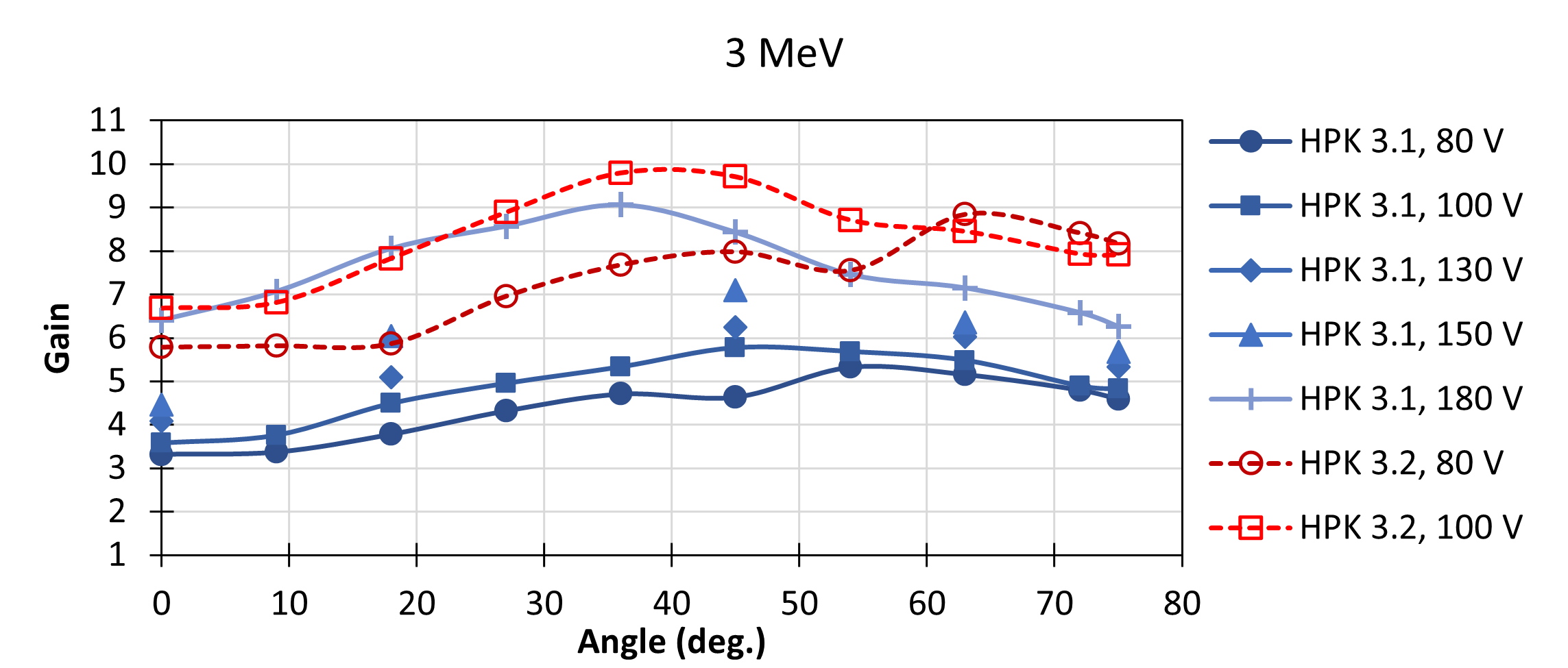}
        \caption{}
        \label{fig:HPK3p1_3MeV}
    \end{subfigure}
    \caption{Gain as a function of angle for different bias voltages for HPK~3.1 and 3.2 LGADs, for proton energies of (a) 1.8 MeV and (b) 3 MeV.} 
    \label{fig:gain1}
\end{figure}

The behavior of the devices was probed as a function of the proton incidence angle and applied bias voltage. 
The PIN sensor was tested for only two voltages: 30V to test the response with non-saturated drift velocity and 200V to test the response with saturated drift velocity. 
Almost no variation in the total charge deposition was observed.
The gain as a function of angle for the two LGAD devices for a beam energy of 1.8~MeV and 3~MeV is shown in Fig.~\ref{fig:gain1} (a) and (b), respectively.
The gain increases with the angle of incidence up to 45-50 degrees; then, it decreases to the largest angle (75 degrees).
The electrons drifting toward the gain layer are a projection of the entry track; for a zero-degree angle, the charge density on the gain layer is high but is reduced for angled tracks, so the suppression mechanism is less pronounced.

Geometric considerations can explain the high (over 45~degrees) angle behavior, as shown in \cite{Mazza:2023col}. In that study, the depth of X-ray absorption was found to influence the sensor response such that the gain is effectively less for charge depositions closer to the gain layer.
The proton energy deposition profile has the properties of a Bragg peak, depositing a large amount of energy at the stopping point. 
At a large angle, the proton deposits a larger amount of charge closer to the gain layer, given that the track length is always the same for 1.8~MeV at all angles and for 3~MeV above 45~degrees.

\begin{figure}
    \centering
    \begin{subfigure}[b]{0.48\textwidth}
        \includegraphics[width=\textwidth]{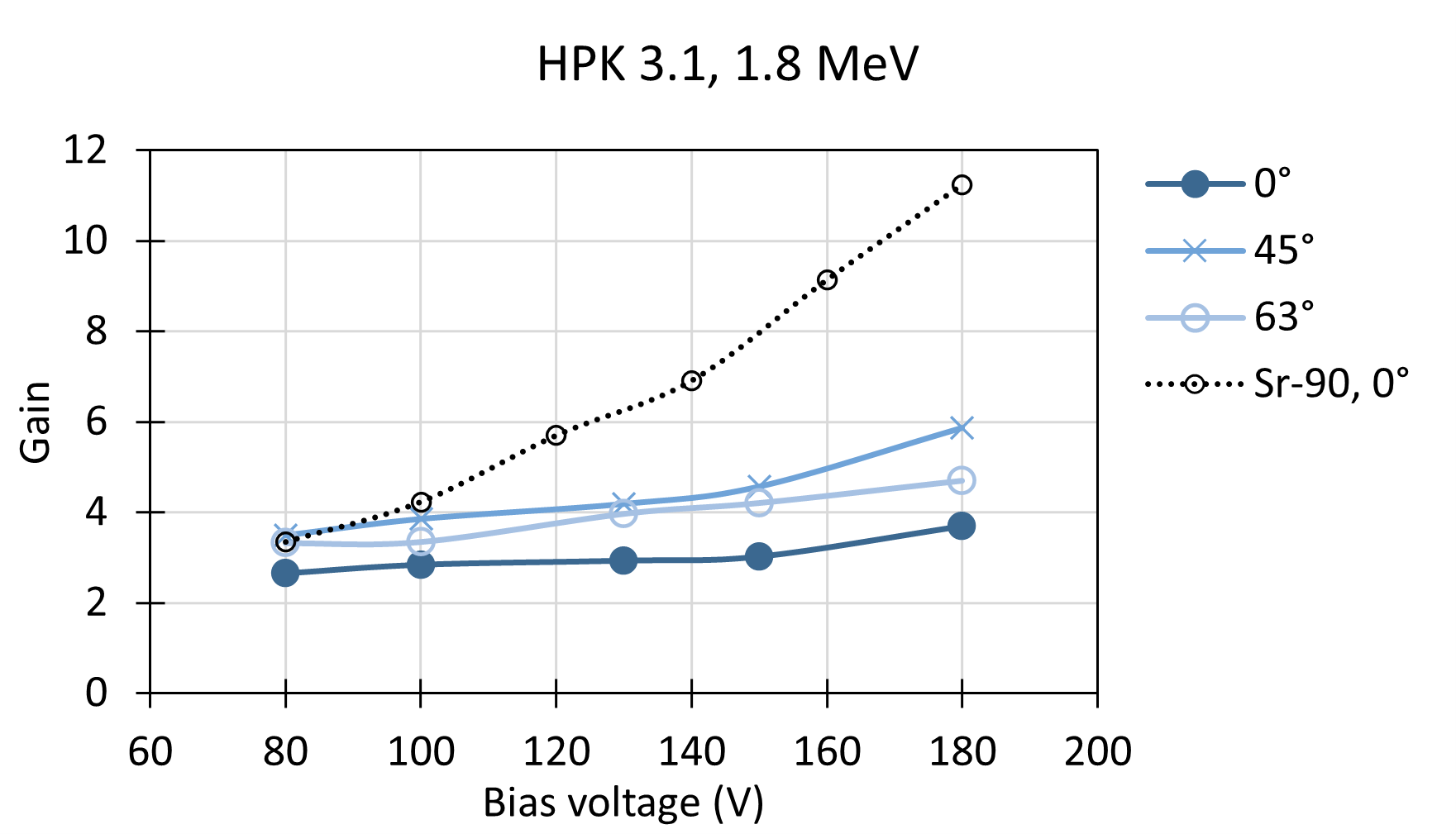}
        \caption{}\label{fig:HPK3p1_1p8MeV}
    \end{subfigure}
    \quad
    \begin{subfigure}[b]{0.48\textwidth}
        \includegraphics[width=\textwidth]{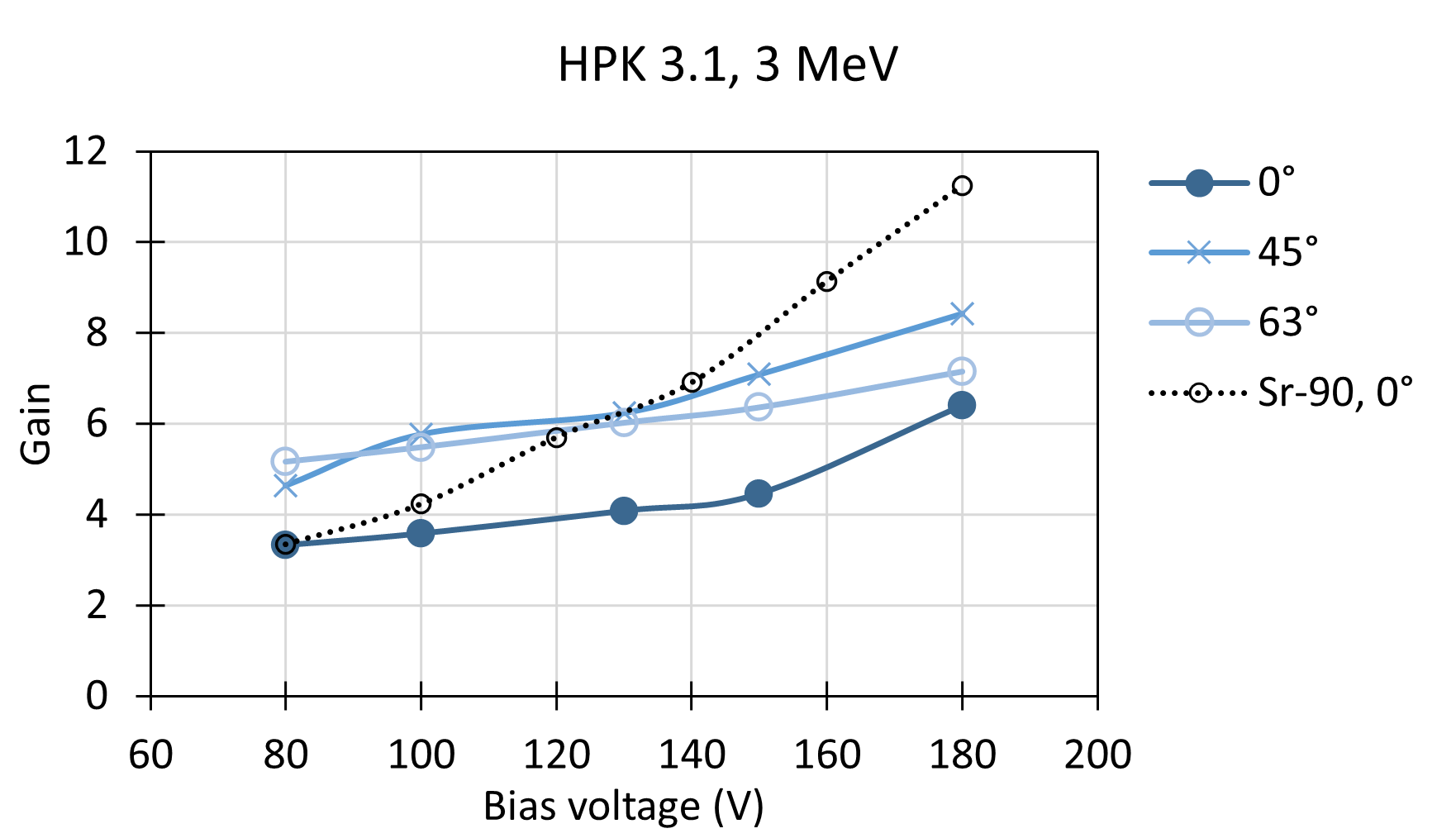}
        \caption{}\label{fig:HPK3p1_3MeV}
    \end{subfigure}
    \caption{Gain of HPK~3.1 LGADs for protons of 1.8~MeV (a) and 3~MeV (b) as a function of bias voltage, compared to the gain determined with Sr-90 beta particles which are MIPs.}
    \label{fig:Gain_BV_HPK3p1_1p8MeV}
\end{figure}

\begin{figure}
    \begin{subfigure}[b]{0.48\textwidth}
        \includegraphics[width=\textwidth]{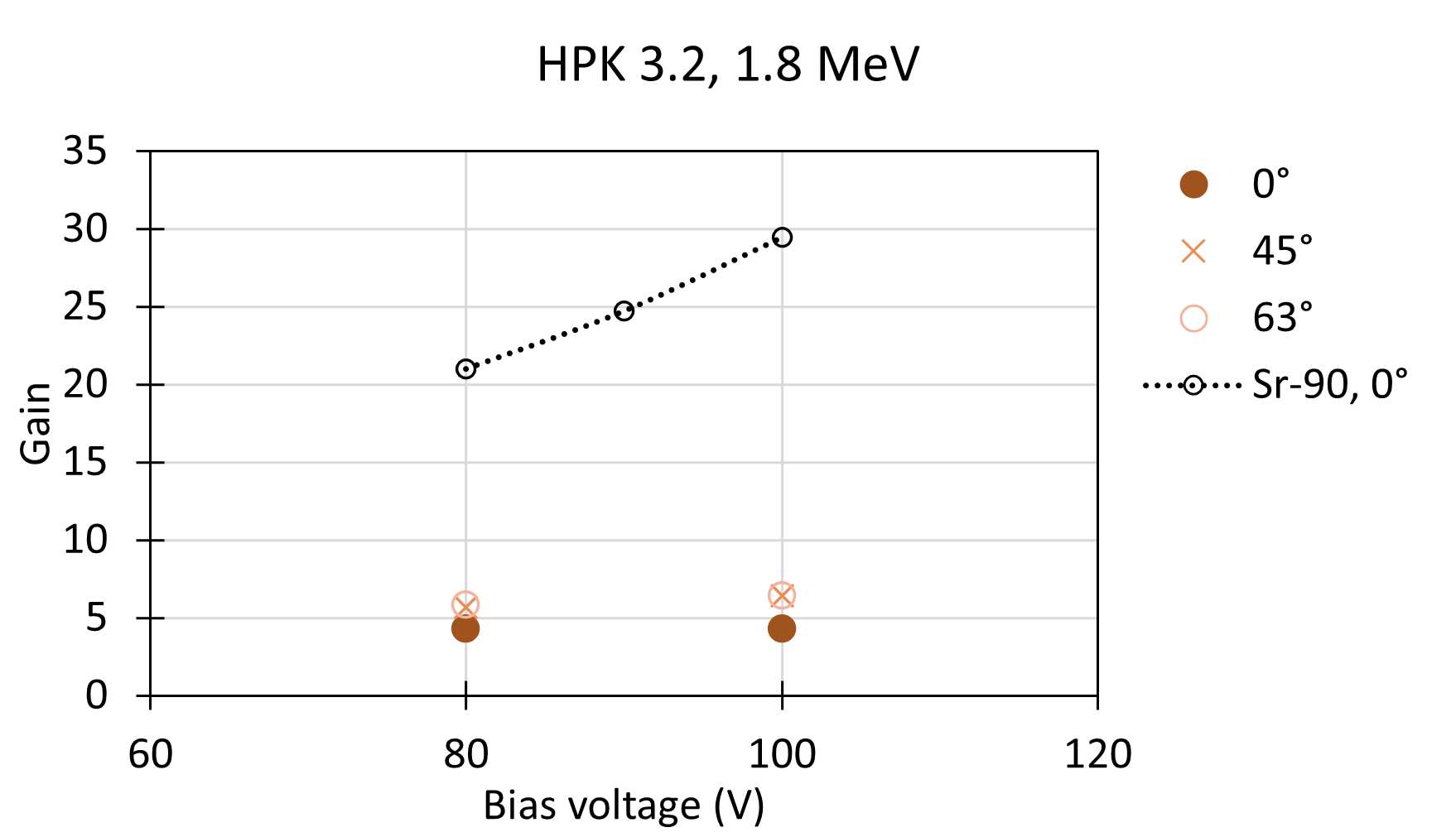}
        \caption{}\label{fig:HPK3p1_1p8MeV}
    \end{subfigure}
    \quad
    \begin{subfigure}[b]{0.48\textwidth}
        \includegraphics[width=\textwidth]{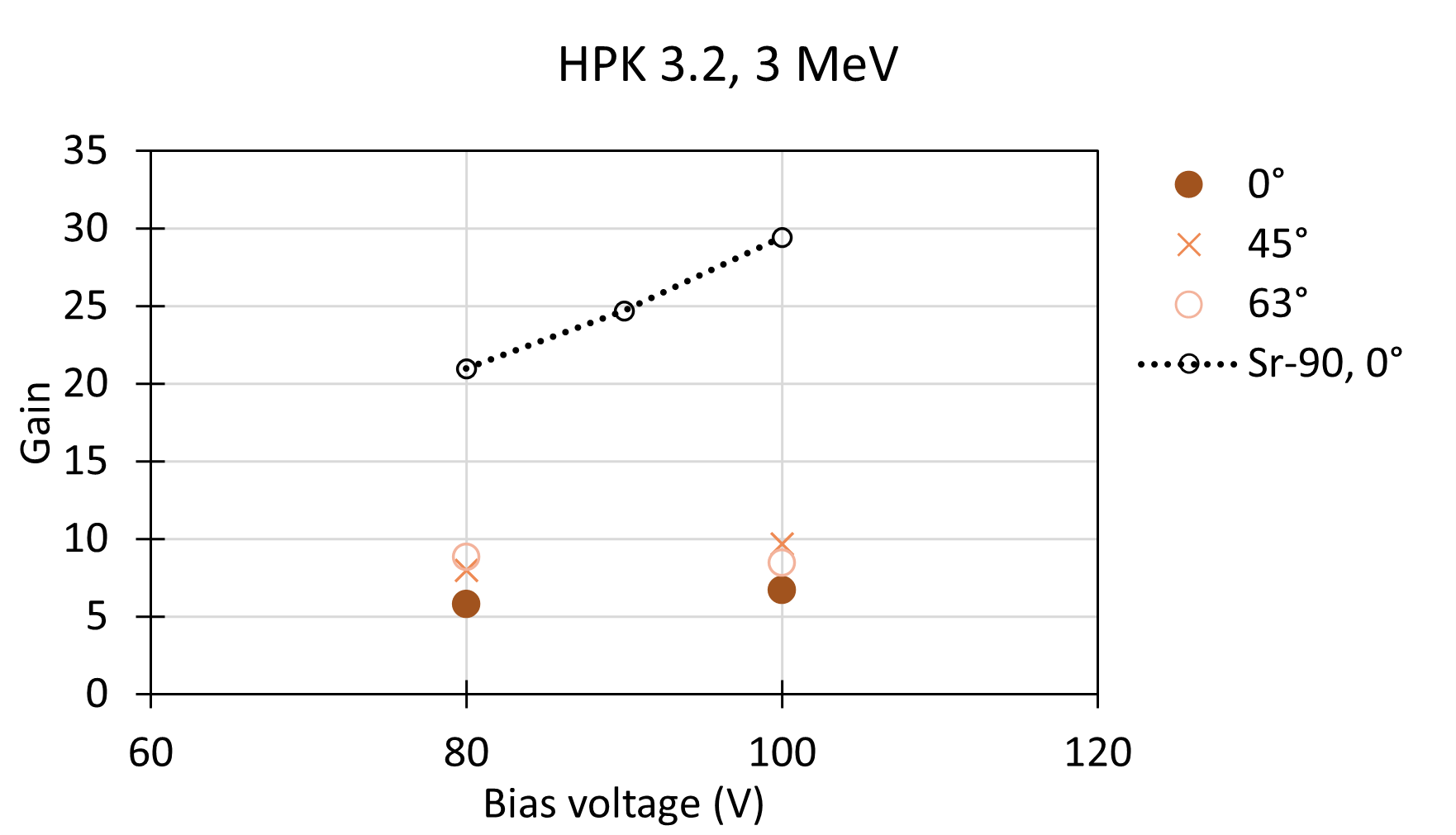}
        \caption{}\label{fig:HPK3p1_3MeV}
    \end{subfigure}

    \caption{Gain of HPK~3.2 LGADs for protons of 1.8~MeV (a) and 3~MeV (b) as a function of bias voltage, compared to the gain determined with Sr-90 beta particles which are MIPs.}
    \label{fig:Gain_BV_HPK3p2_1p8MeV}
\end{figure}

The gain as a function of bias voltage for the two devices is shown in Fig.~\ref{fig:Gain_BV_HPK3p1_1p8MeV} and Fig.~\ref{fig:Gain_BV_HPK3p2_1p8MeV} for selected angles and both proton energies. 
The gain for $^{90}$Sr beta MIPs at zero-degree incidence is shown in the same plot.
The gain for HPK~3.1 is very close between protons and beta particles at low voltage for perpendicular injection. 
In the 3~MeV proton case at low bias voltage, the gain for larger angles is higher than for MIP due to the charge spread across the gain layer.
As the bias voltage increases, so does the initial gain (which is the device gain for charge deposition of 1 MIP); the gain loss is more significant, as expected from the TCAD simulation in Fig.~\ref{fig:alpha_beta}~(b).
The gain loss is always more significant for HPK~3.2, since the initial gain is higher at all the tested voltages.
The results and the high-angle effects observed agree with data for 3~MeV protons presented in~\cite{s22031080}.

\section{Conclusions}

The response to highly ionizing particles of two types of LGAD detectors and one PIN was studied at the CENPA tandem accelerator with 1.8~MeV to 3~MeV protons.
The gain in the LGADs was calculated as a function of proton energy, applied bias voltage, and angle of incidence and compared to the response of the same-size PIN detector.
Both HPK~3.1 and HPK~3.2 show substantial gain suppression, which is more significant for higher applied bias voltage.
HPK~3.2, which has a higher initial gain, shows more absolute suppression.
The behavior changes with the angle of incidence of the beam: the gain increases to 40 to 60 degrees of incidence, but then it is again reduced because of the proton stopping position closer to the gain layer due to the large incidence angle.
Another test beam is foreseen in the near future using devices from other producers with lower gain layer doping concentration to investigate if the suppression can be reduced by operating the devices at lower gain but with enough high voltage to saturate the drift velocity, which depends on the bulk doping and thickness; roughly between 50V and 100V after gain layer depletion.


\section{Acknowledgments}
This work was supported by the United States Department of Energy grant DE-FG02-04ER41286 and DE‐FG02‐97ER41020. We thank the CENPA Technical and Accelerator staff for their assistance in this measurement.
C.~Lansdell received undergraduate researcher support through the NSF REU Program at UW, NSF Grant No 2243362.

\tiny
\bibliography{bib/TechnicalProposal,bib/hpk_fbk_paper,bib/HGTD_TDR,bib/SHIN,bib/nizam,bib/others}

\end{document}